# Observation of a non-Abelian Yang Monopole: From New Chern Numbers to a Topological Transition

S. Sugawa, F. Salces-Carcoba, A. R. Perry, Y. Yue and I. B. Spielman

*Joint Quantum Institute, National Institute of Standards and Technology and the University of Maryland, Gaithersburg, MD 20899-8424 USA*

**Because global topological properties are robust against local perturbations, understanding and manipulating the topological properties of physical systems is essential in advancing quantum science and technology. For quantum computation, topologically protected qubit operations can increase computational robustness, and for metrology the quantized Hall effect directly defines the von Klitzing constant. Fundamentally, topological order is generated by singularities called topological defects in extended spaces, and is quantified in terms of Chern numbers, each of which measures different sorts of fields traversing surfaces enclosing these topological singularities. Here, inspired by high energy theories, we describe our synthesis and characterization of a singularity present in non-Abelian gauge theories - a Yang monopole - using atomic Bose-Einstein condensates in a five-dimensional space, and quantify the monopole in terms of Chern numbers measured on enclosing manifolds. While the well-known 1st Chern number vanished, the 2nd Chern number, measured for the first time in any physical settings, did not. By displacing the manifold, we then observed a phase transition from "topological" to "trivial" as the monopole left the manifold.**

60 years ago, Chen-Ning Yang and Robert Mills proposed a non-Abelian gauge field theory, now known as the Yang-Mills theory, to describe elementary particles[1]. The Yang-Mills theory is an important model that includes a higher gauge symmetry than quantum electrodynamics and now forms a cornerstone of standard model physics[2]. In Yang-Mills theory, soliton solutions including monopoles and instantons play a key role, theoretically describing phenomena in high-energy physics and are even predicted to exist in nature[3]. The monopole solutions are sources of non-Abelian gauge fields and give rise to a non-trivial topology.

The physical significance of magnetic monopoles was first addressed by P. A. M. Dirac[4]. Dirac considered the Aharonov-Bohm phase acquired by an electron with charge $q_e$ moving around a magnetic monopole and showed that the monopole charge must be $q_m = nh/q_e$, where $n$ is an integer and $h$ is Planck's constant. Along with this quantization condition, Gauss' law for the magnetic field $\boldsymbol{B}$ must take a quantized value $nh/q_e = \int_{S_2} \boldsymbol{B} \cdot \mathbf{dS}$, that essentially counts the number of magnetic charges inside the manifold $S_2$ – here a two-dimensional surface. The integral is topologically robust against deformation of the enclosing manifold as long as the number of monopoles enclosed is unchanged. The field from Dirac monopoles has been observed in a range of physical systems and the associated topological charge - the 1st Chern number, often referred to as "the Chern number" - has been measured[5,6,7]. In quantum mechanical systems, gauge fields such as the electromagnetic vector potential $\boldsymbol{A}$ take central stage, (recall that in classical electromagnetism $\boldsymbol{B} = \nabla \times \boldsymbol{A}$) and are required to understand nature at the most fundamental level[8]. 50 years after Dirac's discovery, C.-N. Yang found a non-Abelian extension of Dirac's magnetic monopole in Yang-Mills theory[9].

In this Article, we report on the creation of a Yang monopole in a five-dimensional parameter space built from an atomic quantum gases' internal states, and the measurement of its topological charges by characterizing any associated Abelian or non-Abelian gauge field strength, often called curvatures. In order to measure the higher Chern numbers that result from non-Abelian gauge fields, we developed a method to evaluate the local non-Abelian Berry curvatures through the system's non-adiabatic responses. We identify the Yang monopole by measuring both the 1$^\text{st}$ and the 2$^\text{nd}$ Chern number on enclosing manifolds and found that it is the 2$^\text{nd}$ Chern number that quantifies its non-trivial topology. By moving the manifold, we observed a topological transition where the manifold's topology went from "topological" to "trivial" as the monopole exited the manifold, experimentally confirming that the monopole was the source of the non-Abelian field.

**Monopole's gauge field and Chern numbers**

A vector gauge field $\boldsymbol{A}(\boldsymbol{r}) = (A_1, A_2, \ldots, A_n)$ is non-Abelian when the vector components $A_\mu(\boldsymbol{r})$ fail to commute, i.e., $[A_\mu, A_\nu] \neq 0$. The resulting curvature

$$F_{\mu\nu}(\boldsymbol{r}) = \frac{\partial A_\nu}{\partial r_\mu} - \frac{\partial A_\mu}{\partial r_\nu} + i[A_\mu, A_\nu], \qquad (1)$$

is analogues to the magnetic field, and indeed in three spatial dimensions the components of the magnetic field vector $B_\mu = \epsilon_{\mu\nu\lambda} F_{\nu\lambda}/2$ are just elements of the $F_{\nu\lambda}$ matrices ($\epsilon_{\mu\nu\lambda}$ is the rank-3 Levi-Civita symbol and we used Einstein's implied summation convention for repeated indices). In this language the 1$^\text{st}$ Chern number is the integral

$$C_1 = \frac{1}{2\pi} \int_{S_2} \boldsymbol{B} \cdot \mathbf{dS} = \frac{1}{4\pi} \int_{S_2} F_{\mu\nu}\, dr_\mu \wedge dr_\nu, \qquad (2)$$

of the Abelian field strength $\boldsymbol{B}$ (in suitable dimensions) over a closed two-dimensional manifold $S_2$, where $\wedge$ is the wedge product. The general $n$-th Chern number is obtained by replacing the integrand in Eq. (2) with the $n$-th Chern form, a gauge invariant quantity defined by a trace of $n$ wedge products of the curvatures ($F \wedge F \cdots \wedge F$) integrated over a $2n$-dimensional manifold $S_{2n}$[10] (See Methods).

Chern numbers provide a topological classification of monopoles. Monopoles are generally associated with a divergence in the field strength and can contribute a unit of flux through any enclosing manifold. This generalized flux is quantized, and is given by the Chern numbers. In particular, for Yang monopoles, the 1$^{st}$ Chern number is zero, but the 2$^{nd}$ Chern number is either $+1$ or $-1$.

Quantum systems such as ours are described by a Hamiltonian $\widehat{H}(\boldsymbol{r})$ that depends on the generalized "position" $\boldsymbol{r}$ in parameter space. At each position, the system is characterized by the energies $E_n(\boldsymbol{r})$ and the eigenstates $|n(\boldsymbol{r})\rangle$, where $n$ identifies the eigenstate. A gauge potential called the non-Abelian Berry connection $A_\mu^{mn}(\boldsymbol{r}) = i\langle m(\boldsymbol{r})| \partial/\partial r_\mu |n(\boldsymbol{r})\rangle$ is encoded in the system's wave functions, thus for any position $\boldsymbol{r}$, each vector component $A_\mu$ is matrix with indices $m$ and $n$.

As a consequence of these gauge fields, an initial quantum state $|\psi(\boldsymbol{r})\rangle$ can acquire a geometric phase as the location in parameter space is adiabatically changed. For non-degenerate quantum systems, the resulting geometric phase $\Phi = \oint d\boldsymbol{l} \cdot \boldsymbol{A}$ is called the Berry phase[11]. As F. Wilczek and A. Zee showed, a quantum state evolving within a degenerate subspace can acquire a Wilczek-Zee geometric phase, a matrix valued generalization of the Berry phase obtained as the path-ordered line-integral of a non-Abelian gauge potential[12].

## Topological Hamiltonian of our quantum system

We realized a non-Abelian gauge field by cyclically coupling four levels within the hyperfine ground states of rubidium-87 using radio-frequency (rf) and microwave fields (Fig.2a, b), essentially forming a square plaquette. The four couplings were parameterized by two Rabi frequencies $\Omega_A$ and $\Omega_B$, and two phases $\phi_A$ and $\phi_B$ arranged so that the sum of the phases around the plaquette was π. As shown in Fig. 2a and b, this configuration of control fields, along with a detuning $\delta$, gave us an experimentally controllable five-dimensional parameter space labeled by the Cartesian coordinates $\boldsymbol{r} = (-\Omega_B \cos\phi_B, -\Omega_A \cos\phi_A, -\Omega_A \sin\phi_A, \delta, -\Omega_B \sin\phi_B)$. In much the same way that a two-level atom in a magnetic field can be understood in terms of three Pauli matrices $\hat{\sigma}_i$ ($i = x, y, z$), our four-level system is governed by the Hamiltonian

$$\widehat{H} = -\frac{\hbar}{2}\boldsymbol{r}\cdot\widehat{\boldsymbol{\Gamma}} = -\frac{\hbar}{2}(r_1\hat{\Gamma}_1 + r_2\hat{\Gamma}_2 + r_3\hat{\Gamma}_3 + r_4\hat{\Gamma}_4 + r_5\hat{\Gamma}_5), \qquad (3)$$

where $\hat{\Gamma}_i$ ($i$=1,2,…,5) are the 4-by-4 Dirac matrices and $\hat{I}_0$ is the identity matrix. Furthermore, because each of the Dirac matrices commutes with the time-reversal operator $\hat{T}$, the system has time-reversal symmetry (TRS, See Methods). Kramers theorem then implies that the system has two pairs of degenerate energy states[13] (here with energies $E_\pm = \pm\hbar|\boldsymbol{r}|/2$). Thus, each energy (labeled by + or −) has two independent eigenstates $|\uparrow^\pm(\boldsymbol{r})\rangle$ and $|\downarrow^\pm(\boldsymbol{r})\rangle$, each of these pairs define a degenerate subspace (DS). As shown in Fig. 2b, these DS's are characterized by a generalized magnetization vector $\langle \boldsymbol{\Gamma} \rangle = (\langle \hat{\Gamma}_1 \rangle, \langle \hat{\Gamma}_2 \rangle, \langle \hat{\Gamma}_3 \rangle, \langle \hat{\Gamma}_4 \rangle, \langle \hat{\Gamma}_5 \rangle)$ pointing on a unit 4-sphere in our 5-dimensional space, and different configurations within each DS share the same magnetization vector can be pictured in terms of an additional Bloch sphere (green sphere in Fig. 1). A topological singularity exists at $\boldsymbol{r} = 0$, and we will show by direct measurement that this is a Yang monopole.

## Quantum control and measurement

We began by demonstrating the control and the measurement capabilities of our system. We first prepared the system in its ground state at the position $\boldsymbol{r}_0 = r_0(-1, -1, 0, 0, 0)/\sqrt{2}$ in parameter space, where the generalized magnetization is $\langle \boldsymbol{\Gamma} \rangle = (-1, -1, 0, 0, 0)/\sqrt{2}$. Then, by ramping $\phi_A$, we slowly moved the system around the circle $\boldsymbol{r}(t) = r_0(-1, -\cos(2\pi t/T), -\sin(2\pi t/T), 0, 0)/\sqrt{2}$ shown in Fig. 3a, where $T$ is the full ramp time, and $r_0 = |\boldsymbol{r}_0| = 2\pi \times 2$ kHz.

We measured the state as it evolved within the DS for this nearly adiabatic ramp (Fig. 3b), and identified the orientation within the DS by performing quantum state tomography, giving the expectation values of the Pauli matrices in the ground DS (see Methods). As seen in Fig. 3b, after the control field completed one cycle, the orientation of the state vector within the DS differed from its initial value. Geometric phases can describe all consequence of adiabatic motion, and after one cycle, the Berry's phase from an Abelian gauge field would give only an overall phase, leaving the state vector otherwise unchanged. In agreement with our numerical simulation (curves in Fig. 3b), this shows that the observed evolution resulted instead from the Wilczek-Zee phase derived from a non-Abelian gauge field.

We then measured $\langle \hat{\Gamma}_4 \rangle$ during this ramp, and noted a small deflection of the state vector's magnetization due to remnant non-adiabatic effects (Fig. 3c). In linear response theory, deviations from adiabaticity can be described in terms of the state vector's response to a generalized force $M_\mu = -\langle \partial H / \partial r_\mu \rangle$ acting on the state (Fig.1b). For a conventional Abelian system, the force

$$M_\mu = v_\lambda F_{\mu\lambda}, \qquad (4)$$

resulting from a parameters $r_\lambda$ changing with velocity $v_\lambda$ is analogues to the Lorentz force[14,15]. This relation gives the driving force behind the topological and geometrical charge pumps recently realized in ultracold atoms[16,17,18]. In both crystalline and optical lattices, the same relation underlies in the anomalous quantum Hall effect[19,20,21].

Owning to the system's symmetry, the generalized geometric force from Eq. (4) is constant for our trajectory, inconsistent with the sign change present in the observed deflection (Fig. 3c). To account for this discrepancy, we extended Eq. (4) to accommodate non-Abelian gauge fields, giving the generalized geometric force

$$M_\mu = v_\lambda \mathrm{tr}(\hat{\rho} \hat{F}_{\mu\lambda}) \qquad (5)$$

acting on the state. In contrast to the Abelian case, where the generalized geometric force is simply the product of the local Berry curvature and the velocity, the force in Eq. (5) also depends on the quantum state as expressed by the projected density operator $\hat{\rho}$ within the DS. As we saw, even for adiabatic motion Wilczek-Zee phases can lead to significant evolution within the DS, making Eq. (5) essential for describing generalized geometric forces. Equation (5) is our main theoretical result that we explore experimentally.

The deflection's sign-change we noted in Fig. 3c is now explained by the dependence of the geometric force on the state as it evolved within the DS (Fig. 3c). Indeed the solid curves depict the prediction of this theory and confirm that the geometric force in our experiment cannot be derived from an Abelian gauge potential.

In general, we can observe the full magnetization of the state vector by carefully measuring the expectation values of all the five gamma matrices (Fig. 4b). To demonstrate this capability, we moved along the circle $\boldsymbol{r}(t) = r_0(-\cos(2\pi t/T), -\cos(2\pi t/T), -\sin(2\pi t/T), 0, \sin(2\pi t/T))/\sqrt{2}$ shown in Fig. 4a, and obtained $\langle \boldsymbol{\Gamma}(t) \rangle$. Figure 4b shows that $\langle \widehat{\boldsymbol{\Gamma}} \rangle$ nearly followed the adiabatic trajectory (red curves), always oriented parallel to $\boldsymbol{r}$, but was slightly deflected due to the non-adiabaticity (theory shown by black curves).

## Non-Abelian Berry curvatures and Chern numbers

With the ultimate goal of evaluating Chern numbers in mind, we characterized the non-Abelian Berry curvatures on spherical manifolds in parameter space. Accordingly, we adopt spherical coordinates described by a radius $r$ and four angles $(\theta_1, \theta_2, \phi_1, \phi_2)$, that are related to our experimental control parameter space via $\Omega_A = r \sin\theta_1 \cos\theta_2$, $\Omega_B = r \sin\theta_1 \sin\theta_2$, $\delta = r \cos\theta_1$, $\phi_1 = (\phi_A + \phi_B)/2$ and $\phi_2 = (\phi_A - \phi_B)/2$.

After preparing the system in its ground state at $\boldsymbol{r}_0$, we measured the deflection along the $\theta_1$-direction, while rotating the control field along $\boldsymbol{r}_\pm(t) = r_0(-\cos(2\pi t/T), -\cos(2\pi t/T), \mp\sin(2\pi t/T), 0, \mp\sin(2\pi t/T))/\sqrt{2}$ by ramping $\phi_1$ from 0 to $\pm\pi$ (half-circles in Fig. 5a). The geometric force $M_{\theta_1}$ is directly obtained from the deflection of $\langle \hat{\Gamma}_4 \rangle$. Figure 5b plots the deflection during this ramp for four different initial states (marked by A-D in Fig. 5d) within the DS, manifesting the state dependence of the geometric force in the non-Abelian gauge field in contrast to Abelian cases. The net deflection during any given ramp gives the integrated geometric force.

To confirm that our drive was in the linear response regime, we measured the geometric force as a function of ramp time $T$ (Fig 5c). From both the data and theory (dashed curves), the geometric force (solid curves) is linear with respect to velocity for $T \geq 12\pi/r$.

The components of the Berry curvatures can be reconstructed from the integrated geometric force. Due to the system's symmetry, the geometric force was constant during the ramp. By measuring the geometric force experienced by four independent initial states all within the DS, we determined the four independent parameters present in the the 2-by-2 matrices describing each element (labeled by *m* and *n*) of the non-Abelian Berry curvature $F_{\mu\lambda}^{m,n}$. Following this procedure for $T \geq 12\pi/r$, we obtained $2r_0^2 \hat{F}_{\phi_1\theta_1} = 0.01(3)\hat{I}_0 + (-0.06(5), 0.08(5), 0.98(3))\cdot \hat{\boldsymbol{\sigma}}$, in agreement with the theoretical value, $2r_0^2 \hat{F}_{\phi_1\theta_1} = \hat{\sigma}_z$.

We thoroughly investigated the state-dependence of the geometric force by studying the evolution of 225 initial states covering the Bloch sphere of the initial DS (Fig. 5d and Methods). For each initial state, we recorded the deflection after a 250 µs ramp to obtain the Berry curvature component $\mathrm{tr}(\hat{\rho}\hat{F}_{\phi_1\theta_1})$. Figure 5d shows the initial-state

Bloch sphere colored according to the curvature; the theoretically computed result (top) is in good agreement with experiment (bottom).

By changing the trace and the direction along which we measure the deflection, other components of the curvatures can be measured. For example, by rotating the control field along $\boldsymbol{r}_{\pm}(t) = r_0(-\cos(2\pi t/T), -\cos(2\pi t/T), \mp\sin(2\pi t/T), 0, \pm\sin(2\pi t/T))/\sqrt{2}$ by ramping $\phi_2$ and measuring the deflection along the $\theta_2$ direction, we obtained $2r_0^2 \hat{F}_{\phi_2\theta_2} = -0.08(3)\hat{I}_0 + (-0.12(5), -0.07(5), 1.00(3)) \cdot \hat{\boldsymbol{\sigma}}$, also in good agreement with the theoretical value, $2r_0^2 \hat{F}_{\phi_2\theta_2} = \hat{\sigma}_z$.

Just as for classical electromagnetism, where the fields from electric or magnetic sources fall off as $1/r^2$, the non-Abelian gauge field strength also follows a $1/r^2$ scaling law as required by the generalized Gauss's law [Eq. (2)] that defines the 2nd Chern number. By repeating the same Berry curvature measurement ($\hat{F}_{\phi_2\theta_2}$) for different $r_0$ (keeping $2\pi/rT = 0.25$ constant to remain in the linear response regime), the Berry curvature components $\hat{F}_{\phi_2\theta_2}$ indeed had the $1/r^2$ scaling of a monopole source (Fig. 5e); this also implies that $\hat{F}_{\phi_2\theta_2}$ diverges at $r \to 0$.

Taken together, these fields provide sufficient information to extract the 2nd Chern number of a 4-sphere with radius $r_0$. We evaluate the 2nd Chern number using the relation

$$C_2 = \frac{3r_0^4}{4\pi^2} \int_{S_4} \text{tr}[F_{\phi_1\theta_1} F_{\phi_2\theta_2}] \, d^4S, \tag{6}$$

where $S_4$ defines the 4-sphere and $d^4S = \sin^3\theta_1 \sin 2\theta_2 \, d\theta_1 d\theta_2 d\phi_1 d\phi_2$. Equation (6) relies on our Hamiltonian's rotational symmetry, that gives the numerically confirmed relations $\mathrm{tr}\left[F_{\phi_1\theta_1}F_{\phi_2\theta_2}\right] = \mathrm{tr}[F_{\phi_1\theta_2}F_{\theta_1\phi_2}] = \mathrm{tr}\left[F_{\phi_1\phi_2}F_{\theta_2\theta_1}\right]$. From the non-Abelian Berry curvature measurements in the previous section, we directly obtained $C_2 = 2r_0^4\,\mathrm{tr}\left[F_{\phi_1\theta_1}(\boldsymbol{r}_0)F_{\phi_2\theta_2}(\boldsymbol{r}_0)\right] = 0.97(6)$ for the ground state, consistent with the theoretical value $C_2 = 1$. We repeated the measurements for the excited state and found $C_2 = 2r_0^4\,\mathrm{tr}\left[F_{\phi_1\theta_2}(\boldsymbol{r}_0)F_{\theta_1\phi_2}(\boldsymbol{r}_0)\right] = -0.93(6)$ also in agreement with the theoretical value $C_2 = -1$. These non-zero Chern numbers inform us that the manifold is "topological".

Due to our system's the TRS, the 1$^\mathrm{st}$ Chern form is zero, and therefore Eq. (2) for the first Chern number should be zero for both degenerate manifolds. Indeed, all the measured non-Abelian Berry curvatures were traceless ($r_0^2\mathrm{tr}[\hat{F}_{\phi_1\theta_1}] = -0.04(2)$ and $r_0^2\,\mathrm{tr}[\hat{F}_{\phi_2\theta_2}] = 0.01(2)$ for the ground state, and $r_0^2\,\mathrm{tr}[\hat{F}_{\phi_1\theta_2}] = -0.02(3)$ and $r_0^2\,\mathrm{tr}[\hat{F}_{\phi_2\theta_1}] = 0.00(3)$ for the excited state), so that the 1$^\mathrm{st}$ Chern number, which is the surface integral of the trace of the individual curvatures, were also zero. Thus, the non-trivial topology of the monopole field is not expressed by a 1$^\mathrm{st}$ Chern number.

**Topological Transition**

We concluded our measurements of the system's topology by inducing a "topological" to "trivial" phase transition by displacing the 4-sphere in parameter space from the origin by an amount $r_\mathrm{offset}$ (Fig. 6a). The topological transition occurs at the critical displacement $r_\mathrm{crit} = r_0$ when the Yang monopole departs the manifold. Figure 6b shows our observed transition of the 2$^\mathrm{nd}$ Chern number from $\pm 1$, for the ground and excited states, to zero as the offset coupling $r_\mathrm{offset}$ was increased. This transition is associated with the topology of the manifold changing from "topological" to "trivial". The smoothness of the observed transition was due to the breakdown of the linear response near the transition point. Our theory (continuous curves in Fig. 6b) shows that slower ramps enlarge the region in which linear response is valid and make the transition sharper (Fig. 6b). Topological phase transitions have been observed in a range experiments[6,7,21], however, in all of these cases the observed topological phases were only identified by a Dirac monopoles' $1^{st}$ Chern number, and enclosing 2-dimensional manifolds. However, in our system, the $1^{st}$ Chern number is everywhere zero and the $2^{nd}$ Chern number characterizes the topological phase, arising from a Yang monopole at the origin of parameter space. The opposite topological charges observed in the ground or excited manifolds result from a monopole in one manifold acting as anti-monopole in the other. With these Chern number measurements, we confirmed that the engineered topological singularity in our system was indeed a Yang monopole.

## Discussion and Outlook

We directly measured a higher-order Chern number on a four-dimensional manifold derived from a Yang monopole in a five-dimensional parameter space built from the synthetic dimensions[22,23] given to us by the internal state of atomic Bose-Einstein condensates. This Chern number characterizes a source of gauge field with higher symmetry, a symmetry that naturally arise in particle physics in contexts such as in quantum chromodynamics.

The monopole field and the second Chern number have been discussed theoretically in the context of four-dimensional quantum Hall effect (4DQH)[24,25], spin-Hall effect[26], exotic charge pumping[27] and fermionic pairing[28] in condensed matter systems. The topological model we explored experimentally is equivalent to the (4+1)-D lattice Dirac Hamiltonian relevant to 4DQH. The 4DQH is a generalized quantum Hall effect, and is the root state of a family of topological insulators[29]. A conformal map from a four-dimensional spherical manifold in parameter space to a four-dimensional crystal momentum space, 4-torus, directly recasts our Hamiltonian to the Dirac Hamiltonian.

Our work lays the groundwork for simulating objects in high-energy physics with atomic quantum systems. Lattice extensions of our work, where lattice sites or bands play the role of spin states, may allow quantum simulation of emergent many-body dynamics with non-Abelian gauge fields with highly controllable ultracold quantum gases systems[30,31,32,33]

**Acknowledgement** This work was partially supported by the ARO's Atomtronics MURI, and by the AFOSR's Quantum Matter MURI, NIST, and NSF through the PFC at the JQI. S.S. acknowledges JSPS (fellowship for research aboard).

**Contributions** S.S. conceived the project and performed the theoretical work. S.S. conducted the experiment and analysed the data. S.S, A.R.P, F. S.-C. and I.B.S. contributed to the rubidium BEC apparatus. All authors substantially participated in the discussion and the writing of the manuscript.

**Corresponding author** Seiji Sugawa (sugawa@umd.edu)

**Competing financial interests** The authors declare no competing financial interests.

**Figures and Figure legends**

**Figure 1**

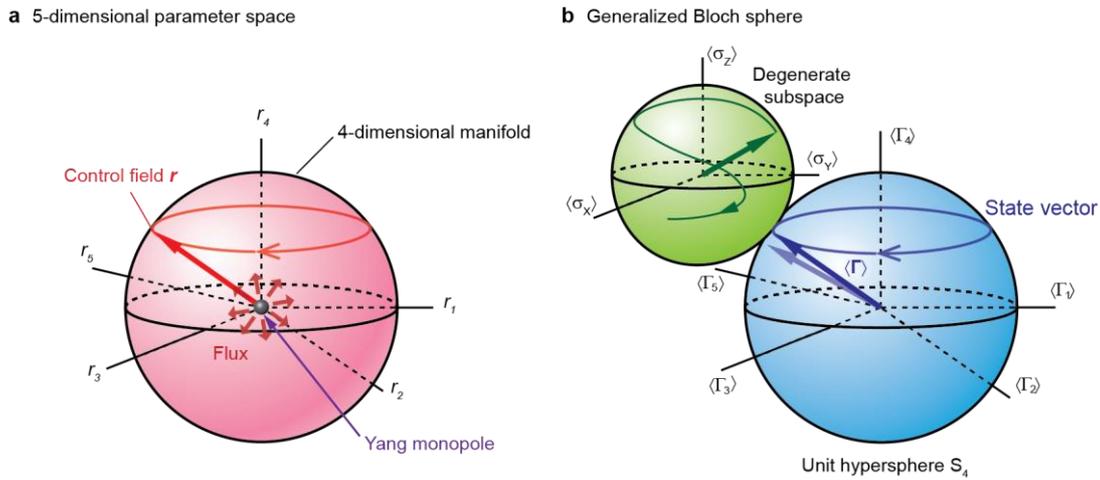

**Figure 1 | Non-Abelian monopole and the appearance of non-trivial topology**

(a) Five-dimensional parameter space. The system has a topological defect at the origin, a Yang monopole, providing a source of non-Abelian gauge field. The topological invariant associated with the monopole is the $2^{nd}$ Chern number, defined on an enclosing four-dimensional manifold.

(b) The system's quantum state can be mapped onto generalized Bloch spheres. An additional Bloch sphere defines the wavefunction within each DS is required to fully define our systems eigenstates. The leading order correction to the adiabaticity changes to $r$, where five-dimensional magnetization vector $\langle \Gamma \rangle$ remains in parallel with $r$, is small deflection in $\langle \Gamma \rangle$.

**Figure 2**

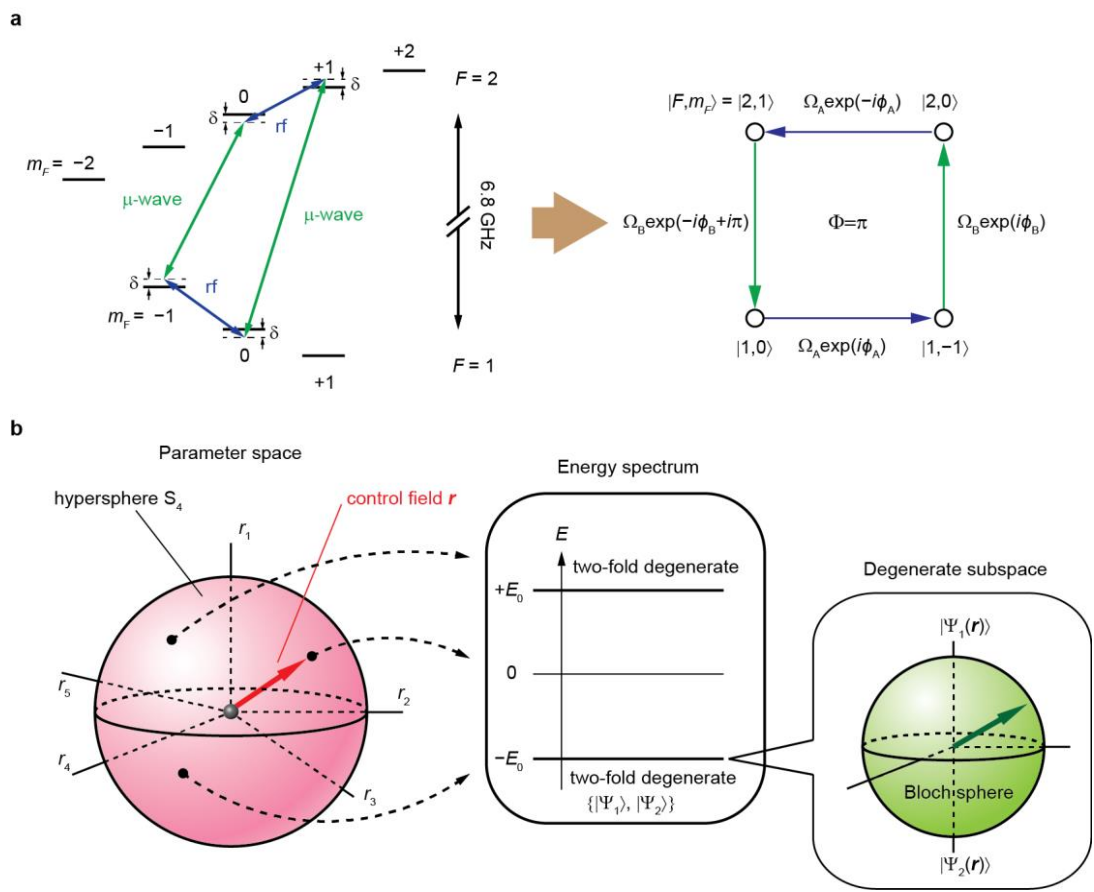

**Figure 2 | Schematic illustration of the experiment**

**(a)** Schematic of our implemented coupling using four hyperfine ground states of rubidium-87. The four states were cyclically coupled with rf and microwave fields. The right panel shows the resulting plaquette and the associated coupling parameters.

**(b)** At any point in the five-dimensional parameter space the energy spectrum forms a pair of two-fold degenerate manifolds with the energy gap equal to $\hbar|\boldsymbol{r}|$. Each degenerate subspace can be represented by a Bloch sphere.

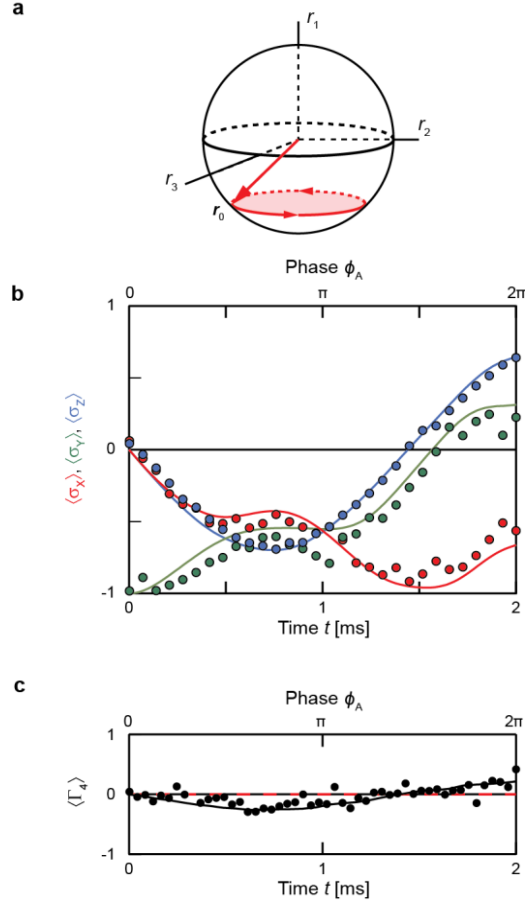

**Figure 3 | State evolution under a non-Abelian gauge field**

**(a)** Schematic of the control field trajectory. The two phases ($\phi_A, \phi_B$) were ramped for $T = 2$ ms with $r/2\pi = 2$ kHz, $\theta_1 = \pi/2$, $\theta_2 = \pi/4$ **(b)** Adiabatic response of pseudo-spin magnetization within the ground DS Bloch sphere, showing the non-trivial acquisition of a Wilczek-Zee phase. **(c)** Deflection during the phase ramp. The state was slightly deflected along $\langle \hat{\Gamma}_4 \rangle$ resulting from our non-infinite ramp time (black symbol). The red dashed lines plots the computed zero value of $\langle \hat{\Gamma}_4 \rangle$ for an adiabatic trajectory, while the black curve was computed using both the finite ramp time and the measured Wilczek-Zee phase.

**Figure 4**

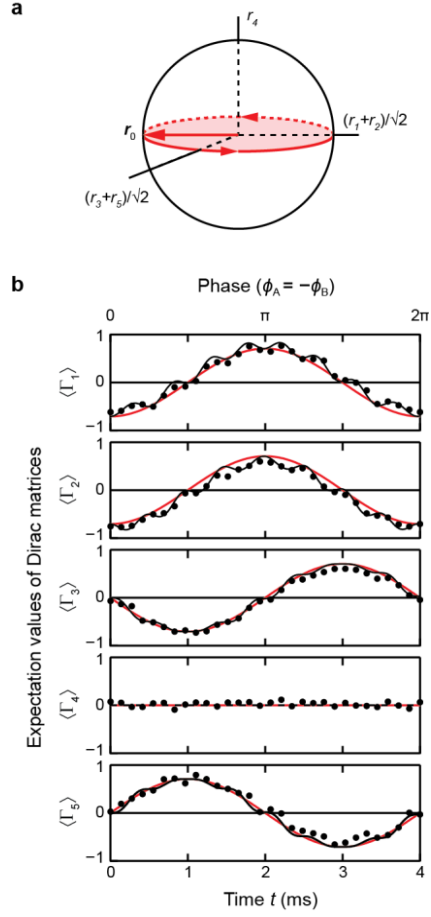

**Figure 4 | Generalized magnetization**

(a) Schematic of the control field trajectory. The two phases ($\phi_A, \phi_B$) are ramped for $T = 4$ ms with $r/2\pi = 2$ kHz, $\theta_1 = \pi/2$, $\theta_2 = \pi/4$, resulting from the laboratory parameters $\Omega_A/2\pi = \Omega_B/2\pi = 1.41$ kHz and $\delta = 0$. (b) The quantum states were measured by evaluating the expectation values of the five Dirac matrices. The red dashed curves plot the trajectory expected for adiabatic motion, while the black curves are numerical simulations including our finite ramp time.

**Figure 5**

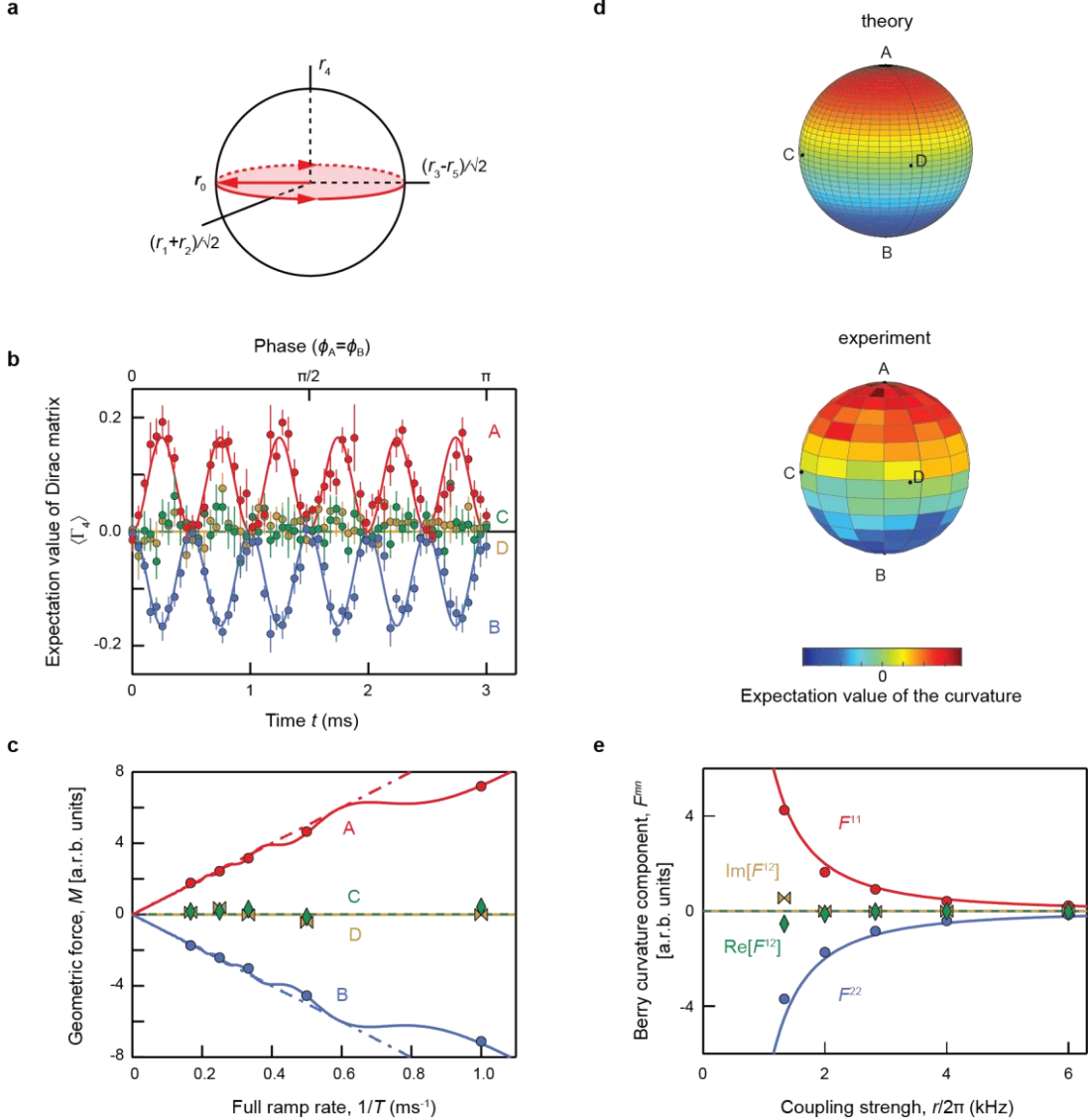

**Figure 5 | Deflection of states within the ground state manifold due to non-Abelian Berry curvatures**

(a) Schematic of the control field trajectory. (b) The state's deflections along $\theta_1$ were measured during the $T = 6$ ms ramp. $\langle \hat{\Gamma}_4 \rangle$ was measured for four independent initial states ($|A\rangle, |B\rangle, |C\rangle, |D\rangle$) within the DS at $r_0$. Here $|A\rangle$ and $|B\rangle$ are the basis states for the DS (See Methods), and $|C\rangle = |A\rangle + |B\rangle$, $|D\rangle = |A\rangle + i|B\rangle$. (c) The linearity of the geometric force with respect to $1/T$ measured for the four initial states ($|A\rangle, |B\rangle, |C\rangle, |D\rangle$). The dashed lines assume linearity and the solid line are the outcome of numerical simulations. (d) The expectation value of the non-Abelian Berry curvature $\text{tr}(\hat{\rho}\hat{F}_{\phi_1\theta_1})$ in the ground state manifold are mapped onto Bloch spheres associated with the state within the DS at $r_0$. The four initial state used in (a), (b) and (c) are also shown in the theory and the experiment plots. **(e)** $1/r^2$ scaling in the strength of the curvature. The matrix components of the curvature $F_{\phi_2\theta_2}^{m,n}$ are evaluated for various $r_0$. The data shows excellent agreement with the theory that exhibits $1/r^2$ dependence (solid lines).

**Figure 6**

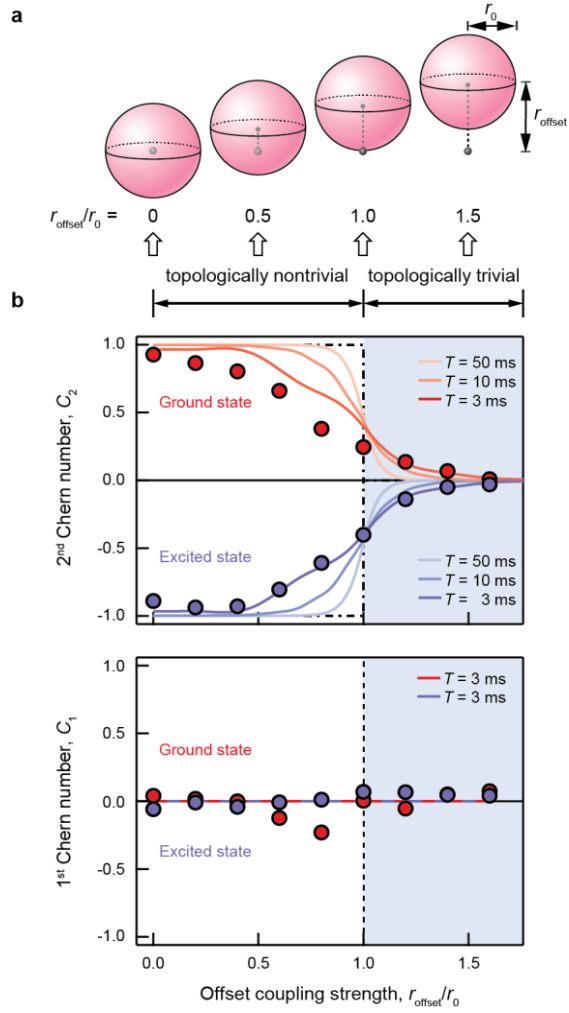

**Figure 6 | Topological transition from a Yang monopole**

**(a)** Schematic illustration of topological transition. Suitable (hyper-) spherical manifolds with radius $r_0$ were offset from the origin by $r_{\text{offset}}$ to evaluate both $C_1$ and $C_2$. At the critical value ($r_{\text{crit}} = r_0$), the monopole exits the manifolds. **(b)** Measured Chern numbers. When the manifold crossed $r_{\text{crit}} = r_0$, $|C_2|$ rapidly decreased from unity to zero (top panel), while $C_1$ was constantly zero for both the ground (red) and the excited (blue) states (bottom panel). Numerical simulations ($T$ =3, 50, 100 ms) and theory curves (dash-dot lines) are also shown. The data were taken for $T$ =3 ms.

## METHODS

### Preparation of Bose-Einstein Condensate in an optical trap

Bose-Einstein Condensate of rubidium-87 of $N \sim 1 \times 10^5$ was prepared in 1064 nm cross optical dipole trap following standard laser cooling technique and evaporative cooling in hybrid magnetic and optical trap. The final trapping frequencies were $2\pi \times (50, 110, 70)$ Hz. The atoms were initially prepared in the $|F, m_F\rangle = |1, -1\rangle$ hyperfine ground state ($F$ is the total angular momentum and $m_F$ is the projection along the quantization axis) at a bias field of 19.8 Gauss before any operation was performed.

### Derivation of the system's Hamiltonian

The Hamiltonian of our four-level quantum system can be written as

$$\widehat{H}(t) = \hbar \begin{pmatrix} E_1/\hbar & \Omega_{12}(t) & 0 & \Omega_{14}(t) \\ \Omega_{12}(t) & E_2/\hbar & \Omega_{23}(t) & 0 \\ 0 & \Omega_{23}(t) & E_3/\hbar & \Omega_{34}(t) \\ \Omega_{14}(t) & 0 & \Omega_{34}(t) & E_4/\hbar \end{pmatrix},$$

where $\Omega_{12}(t) = \Omega_{12}\cos(\omega_{12}t + \phi_{12})$, $\Omega_{23}(t) = \Omega_{23}\cos(\omega_{23}t + \phi_{23})$, $\Omega_{34}(t) = \Omega_{34}\cos(\omega_{34}t + \phi_{34})$, $\Omega_{14}(t) = \Omega_{14}\cos(\omega_{14}t + \phi_{41})$ are the driving fields and $E_i$ is the energy of $i$-th state. Here $\Omega_{ij}$ is the amplitude of the Rabi frequency, $\omega_{ij}$ is the frequency and $\phi_{ij}$ is the phase of the driving field that couple $i$-th and $j$-th states. As a basis, bare hyperfine ground states ($|1\rangle = |1,0\rangle, |2\rangle = |1,-1\rangle, |3\rangle = |2,0\rangle$ and $|4\rangle = |2,1\rangle$) are taken. By choosing $\omega_{12} = (E_2 - E_1)/\hbar + \delta/2$, $\omega_{23} = (E_3 - E_2)/\hbar - \delta/2$, $\omega_{34} = (E_4 - E_3)/\hbar + \delta/2$, $\omega_{14} = (E_4 - E_1)/\hbar - \delta/2$ for the driving frequencies, and $\Omega_{12} = \Omega_{34} = \Omega_A$, $\Omega_{23} = \Omega_{14} = \Omega_B$, $\phi_{12} = -\phi_{34} = \phi_A$, $\phi_{23} = -\phi_{41} + \pi = \phi_B$ for the Rabi frequencies, the Hamiltonian in the interaction picture can be written as in Eq. (3).

**Dirac matrices**

As the representation of the Dirac matrices, we took $\hat{\Gamma}_1 = \hat{\sigma}_y \otimes \hat{\sigma}_y$, $\hat{\Gamma}_2 = \hat{I}_0 \otimes \hat{\sigma}_x$, $\hat{\Gamma}_3 = -\hat{\sigma}_z \otimes \hat{\sigma}_y$, $\hat{\Gamma}_4 = \hat{I}_0 \otimes \hat{\sigma}_z$, $\hat{\Gamma}_5 = \hat{\sigma}_x \otimes \hat{\sigma}_y$. The Dirac matrices satisfy the Clifford algebra, $\{\hat{\Gamma}_i, \hat{\Gamma}_j\} = 2\delta_{ij} I_0^{4\times 4}$, where $\{,\}$ is the anti-commutator, $I_0^{4\times 4}$ is the 4-by-4 identity matrix and $\delta_{ij}$ is the Kronecker delta. Each Dirac matrix has eigenvalues of $\pm 1$ and have two eigenstates for each value.

**Time-reversal symmetry of the system**

The time-reversal operator of the system is $T = \Theta K$, where $\Theta = i\hat{\sigma}_y \otimes \hat{I}_0$ and $K$ is the complex conjugate operator ($K|n\rangle = |n\rangle^*$). Since the time reversal operator commutes with the Dirac matrices, i.e., $[\Gamma, T] = 0$, and therefore with the system's Hamiltonian, the system has time-reversal symmetry. Due to $T^2 = -1$, from the Kramers theorem, a degenerate Kramers pair is formed in the energy spectrum.

**State preparation**

In the generalized Bloch picture of our system shown in Fig. 1, the magnetization vector points at $\langle \boldsymbol{\Gamma}_N \rangle = (0,0,0,1,0)$ when all the atoms were in $|1,-1\rangle$ state, which corresponds to the control field pointing along $\boldsymbol{r}_N = r_0(0,0,0,1,0)$. Due to the two-fold degeneracies, $|\uparrow_- (\boldsymbol{r}_N)\rangle = |1,0\rangle$ and $|\downarrow_- (\boldsymbol{r}_N)\rangle = |2,0\rangle$ are the eigenstates in the ground level and $|\uparrow_+ (\boldsymbol{r}_N)\rangle = |1,-1\rangle$ and $|\downarrow_+ (\boldsymbol{r}_N)\rangle = |2,+1\rangle$ are eigenstates in the excited level. Any eigenstate within the DS at $\langle \boldsymbol{\Gamma}_N \rangle$ is prepared by making the coherent superposition of $|\uparrow_\pm(\boldsymbol{r}_N)\rangle$ and $|\downarrow_\pm (\boldsymbol{r}_N)\rangle$ which takes the form $|\Psi_\pm\rangle = \cos(\Theta/2)|\uparrow_\pm(\boldsymbol{r}_N)\rangle + \exp(i\Phi) \sin(\Theta/2)|\downarrow_\pm(\boldsymbol{r}_N)\rangle$. Here, $\Phi$ is the relative phase

between the two states and $\Theta$ defined the ratio of the population. Experimentally, we either used the clock transition ($|1,0\rangle \leftrightarrow |2,0\rangle$), resonant three-state coupling ($|1,0\rangle \leftrightarrow |1,-1\rangle \leftrightarrow |2,0\rangle$) or the cyclic plaquette coupling to prepare the superposition state in the ground state. To prepare the eigenstates in the excited state, two $\pi$-pulses are applied to the relevant transitions to transfer the coherence to $|\uparrow_+ (\boldsymbol{r}_N)\rangle$ and $|\downarrow_+ (\boldsymbol{r}_N)\rangle$ states. Once an eigenstates at $\boldsymbol{r}_N$ is prepared, the eigenstate at arbitrary $\boldsymbol{r}$ in the parameter space were prepared by pulsing the cyclic coupling with the phases shifted by $\pi/2$ from the target coupling. Specifically, in order to prepare the eigenstate at $\boldsymbol{r} = \boldsymbol{r}_0$, we pulsed a time-independent coupling for a duration $t = \pi/2\Omega$, which can be expresses by a unitary operator $\widehat{U}_{\text{prep}} = \exp(-i\widehat{H}_{\text{prep}}t/\hbar)$ with $\widehat{H}_{\text{prep}} = \hbar\Omega(-\hat{I}_0 \otimes \hat{\sigma}_y + \hat{\sigma}_y \otimes \hat{\sigma}_x)/2$. The unitary operator gives a one-to-one correspondence between the degenerate subspace at $\boldsymbol{r}_N$ and $\boldsymbol{r}_0$. The operation is analogues to rotation of magnetization vector in spin-1/2 system.

**Degenerate subspaces**

A basis for the degenerate subspace can be defined by a unitary operator
$$\widehat{U}_{prep}(\theta_1, \theta_2, \phi_A, \phi_B) = e^{-i\widehat{H}_{\text{prep}}\,\theta_1/\hbar\Omega_0},$$

where

$$\widehat{H}_{\text{prep}} = \frac{\hbar\Omega_0}{2}\begin{pmatrix} 0 & i\cos\theta_2\, e^{i\phi_A} & 0 & -i\sin\theta_2\, e^{i\phi_B} \\ -i\cos\theta_2\, e^{-i\phi_A} & 0 & -i\sin\theta_2\, e^{i\phi_B} & 0 \\ 0 & i\sin\theta_2\, e^{-i\phi_B} & 0 & i\cos\theta_2\, e^{-i\phi_A} \\ i\sin\theta_2\, e^{-i\phi_B} & 0 & -i\cos\theta_2\, e^{i\phi_A} & 0 \end{pmatrix}.$$

For degenerate subspace at $\boldsymbol{r} = \boldsymbol{r}(\theta_1, \theta_2, \phi_A, \phi_B)$, we took the basis as $\widehat{U}|\downarrow_\pm (\boldsymbol{r}_N)\rangle$ for the ground states, for example.

**State evolution within degenerate levels**

In order to measure of state vector within degenerate subspaces, we mapped the state back to the bare states by applying cyclic coupling pulse using the fact that the basis of degenerate subspaces can be related by $\widehat{U}_{prep}$. In the measurement shown in Fig. 3b, for each phase $\phi_A$, the magnetization vector is rotated to $\langle \pmb{\Gamma}_N \rangle$ and the state within the degenerate subspace in evaluated by performing state tomography between the clock states.

**Generalized geometric force**

To derive the generalized geometric force that include the case for non-Abelian gauge field, the Schrödinger equation is solved perturbatively based on the adiabatic perturbation theory for degenerate systems[34], where the wavefunction was expanded in series of ramp velocity. When the terms proportional to the ramp velocity is the only dominant contribution to the generalized force, one obtains the expression in Eq. (5).

**Measuring the generalized magnetization vector**

In order to characterize the quantum state, we measured the expectation values of the Dirac matrices. $\langle \hat{\Gamma}_4 \rangle$ is evaluated from $\langle \hat{\Gamma}_4 \rangle = (N_{2,1} + N_{1,-1} - N_{2,0} - N_{1,0})/(N_{2,1} + N_{1,-1} + N_{2,0} + N_{1,0})$, where $N_{F,m_F}$ is the atom number in $|F, m_F\rangle$. Other $\langle \hat{\Gamma}_i \rangle$ are measured by evaluating $\langle \hat{\Gamma}_4 \rangle$ after applying a unitary operation $U_i = \exp(-i\widehat{H}_{\text{rot},i} t)$ that satisfies $\hat{\Gamma}_i = U_i^\dagger \hat{\Gamma}_4 U_i$. Here, $\widehat{H}_{\text{rot},1} = \hbar\Omega_B \hat{\sigma}_y \otimes \hat{\sigma}_x/2, \widehat{H}_{\text{rot},2} = -\hbar\Omega_A \hat{I}_0 \otimes \hat{\sigma}_y/2, \widehat{H}_{\text{rot},3} = \hbar\Omega_A \hat{\sigma}_z \otimes \hat{\sigma}_x/2, \widehat{H}_{\text{rot},5} = \hbar\Omega_B \hat{\sigma}_x \otimes \hat{\sigma}_x/2$. Experimentally, these operations corresponded to applying $\pi/2$ pulses ($t = \pi/2\Omega_A = \pi/2\Omega_B$) for either the rf or the microwave transitions.

**2nd Chern number for the shifted manifolds**

The 2nd Chern number for the shifted manifolds can be evaluated as

$$C_2 = \frac{3}{2}\Omega_0^4 \, \text{tr}\left[\left(\int \frac{F_{\phi_1\theta_1}(\phi_1)\sin^3\phi_1}{|r(\phi_1)|^2}d\phi_1\right)F_{\phi_2\theta_2}(\phi_1 = 0)\right], \tag{6}$$

which was applied for the excited level case. A different combination of Berry curvatures ($F_{\phi_2\theta_2}, F_{\phi_2\theta_1}$) is chosen for the ground level, which also takes the similar expression. Here we took a coordinate that has a polar axis along $r_0$ and used the $1/|r|^2$ scaling of the Berry curvatures strength. For the other parameters, $\phi_2 = 0$, $\theta_1 = \pi/2$ and $\theta_2 = \pi/4$ are taken and fixed. Although the shifts in the manifold breaking the hyper-spherical symmetry, the azimuthal symmetry around $r_0$ remains. To evaluate the integral in Eq. (6), the control field is rotated as $r(t) = \Omega(t)(-\cos(2\pi vt), -\cos(2\pi vt), \mp\sin(2\pi vt), 0, \mp\sin(2\pi vt))/\sqrt{2}$, where $\Omega(t) = \Omega\sqrt{(1 + 2x_{\text{offset}}\cos(2\pi vt) + x_{\text{offset}}^2)}$ and $x_0 = \Omega/\Omega_{\text{offset}}$.

**Atom number counting**

The atom number is counted by standard absorption imaging technique after time-of-flight of 23.2 ms using $F = 2$ to the excited $F' = 3$ transition. To image atoms in ground $F = 1$ manifold, a short repump laser pulse (resonant to $F = 1$ to the excited $F' = 2$ transition) of 20 μs was applied before the imaging. In order to resolve magnetic sublevels, we apply magnetic field gradient pulse during the TOF to perform Stern-Gerlach measurement. The gradient pulse spatially separates atoms in $|1,0\rangle$ and $|2,0\rangle$ states from those in $|1, +1\rangle$ and $|2, -1\rangle$ states. To image atoms only in the $F = 2$ manifold state, we simply shut the repump pulse during the TOF absorption imaging stage.

**Anti-symmetry in the geometric force**

From the generalized geometric force relation in Eq. (5), when the direction of the ramp is flipped, the sign of the force is also flipped. In the Berry curvature measurements, we take two opposite phase ramps and flipping the sign of the deflection in one of the ramp by also taking into account the system's symmetry. The procedure has the advantage for not requiring the precise knowledge of zero level of the deflection.

**Fluctuation in magnetic fields**

To monitor and compensate the drift in the magnetic field strength, after preparing BEC in $|1,-1\rangle$, we transfer up to 5% of the total atoms to $|2,-2\rangle$ by applying two 400 μs near-resonant microwave pulses with frequency separated by 2.5 kHz. Subsequently after each transfer, atoms in $|2,-2\rangle$ is imaged in-situ to count the transferred atoms numbers. The imbalance of the transferred atoms between the two transfers can be converted to the bias field strength, which was used as an "atomic field monitor" together with a fluxgate field sensor placed near the experiment setup. By compensating slow field drift, the long-term shot-to-shot field fluctuation was suppressed to 50 μG in r.m.s.

**Residual energy gaps**

Since the experimental parameters can slightly deviate from the ones in the system's Hamiltonian, small energy gaps may open between the degenerate energy levels. The residual gaps were measured to be less than 1% of the large energy gap ($\hbar|r|$) by observing coherent oscillation between energy levels, both for the ground and the excited states when $\Omega/2\pi = 2.0$ kHz and $r_{\text{offset}} = 0$. For the short ramp times used in the experiment, both the ground and excited levels can be treated as degenerate levels.